\documentstyle[ltwol,epsfig]{article}

\def\be{\begin{equation}}
\def\ee{\end{equation}}
\def\bea{\begin{eqnarray}}
\def\eea{\end{eqnarray}}

\begin{document}

\title{SUPERSYMMETRY CONTRIBUTION TO BOTTOM QUARK PRODUCTION \\ AT HADRON 
COLLIDERS}

\author{EDMOND~L.~BERGER}
\address{High Energy Physics Division, Argonne National Laboratory, Argonne,
IL 60439, USA\\E-mail: berger@anl.gov}   


\twocolumn[\maketitle\abstracts{A new contribution from pair-production 
of light gluinos, of mass 12 to 16 GeV, with two-body decays into bottom quarks 
and light bottom squarks, helps to obtain a bottom-quark production rate in better 
agreement with hadron collider data.  The masses of the gluino and bottom squark 
are further restricted by the ratio of like-sign to opposite-sign leptons at hadron 
colliders.  Constraints on this scenario from other data are examined and predictions
are made.}]

\section{Motivation and Constraints}
The cross section for bottom-quark production at hadron collider energies exceeds 
the central value of predictions of next-to-leading order (NLO) perturbative quantum 
chromodynamics (QCD) by about a factor of
two~\cite{expxsec}.  This longstanding discrepancy has resisted fully 
satisfactory resolution within the standard model~\cite{qcdrev}.  The NLO 
contributions are large, and it is not 
excluded that further higher-order effects in production and/or fragmentation 
may resolve the discrepancy.  However, the disagreement is surprising because 
the relatively large mass of the bottom quark sets a hard scattering scale at
which fixed-order perturbative QCD computations of other processes are generally
successful.  The data invite the possibility of a contribution from ``new physics".  

In a recent paper~\cite{ourletter}, my collaborators and I explore an explanation 
within the context of the minimal supersymmetric standard model.  We 
postulate the existence of a relatively light gluino $\tilde g$ (mass $\simeq 12$ to 
16 GeV) that decays into a bottom quark $b$ and a light bottom squark $\tilde b$ 
(mass $\simeq 2$ to 5.5 GeV).  The $\tilde g$ and the $\tilde b$ are the spin-1/2 
and spin-0 supersymmetric partners of the gluon ($g$) and bottom quark.  In our 
scenario the masses of all other supersymmetric particles are arbitrarily heavy, 
i.e., of order the electroweak scale or greater.  While the gluino decays 
instantly with 100\% branching fraction into the $b$ and $\tilde b$, the 
$\tilde b$ is either long-lived or decays via R-parity violation into a pair of 
hadronic jets.  We obtain improved agreement with hadron collider rates of 
bottom-quark production.  Several predictions are made that can be 
tested readily with forthcoming data. 

Our assumptions are consistent with all constraints on the masses and
couplings of supersymmetric 
particles~\cite{CHWW,Nappi,CELLO,CLEO,ALEPH}.  If the light bottom 
squark ($\widetilde{b}_1$) is an appropriate mixture of left-handed and 
right-handed bottom squarks, its tree-level 
coupling to the neutral gauge boson $Z$ can be small, leading to good agreement 
with the $Z$-peak observables~\cite{CHWW}.  The couplings 
$Z_{\tilde{b}_1 \tilde{b}_2}$ and $Z_{\tilde{b}_2 \tilde{b}_2}$, where 
$\widetilde{b}_2$ is the heavier bottom squark, survive but present no difficulty 
if $m_{\tilde{b}_2} > 200$ GeV.   Bottom squarks make a small contribution to the
inclusive cross section for $e^+ e^- \rightarrow$ hadrons, in comparison to
the contributions from quark production, and $\tilde{b} \bar{\tilde{b}}$
resonances are likely to be impossible to extract from backgrounds
\cite{Nappi}.  The angular distribution of hadronic jets produced in $e^+ e^-$ 
annihilation can be examined in order to bound the contribution of
scalar-quark production.  Spin-1/2 quarks and spin-0 squarks emerge with
different distributions, $(1 \pm {\rm cos}^2 \theta)$, respectively.  We
re-fit the angular distribution measured by the CELLO
collaboration~\cite{CELLO}, and find it is consistent with the production
of a single pair of charge-1/3 squarks along with five flavors of
quark-antiquark pairs.  A new examination of the angular distribution 
with greater statistics would be valuable.
The exclusion by the CLEO 
collaboration~\cite{CLEO} of a $\tilde b$ with mass 3.5 to 4.5 GeV does not 
apply since that analysis focuses on the decays 
$\tilde b \rightarrow c \em{l} \tilde \nu$ and 
$\tilde b \rightarrow c {\em l}$.
On the other hand, these data might be 
reinterpreted in terms of a bound on the R-parity violating lepton-number 
violating decay of $\tilde b$ into $c {\em l}$.  It would be interesting  
to study the hadronic decays $\tilde b \rightarrow c q$, with $q = d$ or $s$, 
and $\tilde b \rightarrow u s$ with the CLEO data.
In summary, measurements at 
$e^+ e^-$ colliders do not significantly constrain $\tilde b$ masses.  An analysis 
of 2- and 4-jet events by the ALEPH 
collaboration~\cite{ALEPH} disfavors $\tilde g$'s with mass 
$m_{\tilde g} < 6.3$ GeV but not $\tilde g$'s in the mass range of interest to us.  
A renormalization group argument~\cite{Dedes} suggests that the existence of a light 
$\tilde b$ goes hand-in-hand with a comparatively light $\tilde g$ 
($m_{\tilde g} \sim 10$ GeV).   
\section{Differential Cross Section}
Because the excess production rate is observed in all bottom-quark decay
channels, an explanation in terms of new physics is guided towards
hypothesized new particles that decay either like bottom quarks or directly
to bottom quarks.  The former is difficult to implement 
successfully~\cite{ourletter}.  In our scenario, light gluinos are
produced in pairs via standard QCD subprocesses, dominantly $g + g
\rightarrow \tilde g + \tilde g$ at Tevatron and Large Hadron Collider (LHC) 
energies.  The $\tilde g$ has
a strong color coupling to $b$'s and $\tilde b$'s and, as long as its mass
satisfies $m_{\tilde g} > m_b + m_{\tilde b}$, the $\tilde g$ decays
promptly to $b + \tilde b$.  The magnitude of the $b$ cross section, the
shape of the $b$'s transverse momentum $p_{Tb}$ distribution, and the CDF
measurement~\cite{cdfmix} of $B^0 - \bar B^0$ mixing are three features of
the data that help to establish the preferred masses of the $\tilde g$ and
$\tilde b$.

We include contributions from both $q + \bar q \rightarrow \tilde g + \tilde
g$ and $g + g \rightarrow \tilde g + \tilde g$.  The subprocess 
$g + b \rightarrow \tilde{g} + \tilde{b}$ contributes insignificantly.  In 
Fig.~1 we show the integrated $p_{Tb}$ distribution of the $b$ quarks that results
from $\tilde g \rightarrow b + \tilde b$, for $m_{\tilde g} = $14 GeV and
$m_{\tilde b} =$ 3.5 GeV.  The results are compared with the cross section
obtained from next-to-leading order (NLO) perturbative QCD
and CTEQ4M parton distribution functions (PDF's)~\cite{cteq}, with $m_b =$
4.75 GeV, and a renormalization and factorization scale $\mu=\sqrt{m_b^2 +
p_{Tb}^2}$.  SUSY-QCD corrections to $b \bar{b}$ production are not
included as they are not available and are generally expected to be
somewhat smaller than the standard QCD corrections.  A fully differential
NLO calculation of $\tilde g$-pair production and decay does not exist
either.  Therefore, we compute the $\tilde g$-pair cross section from the
leading order (LO) matrix element with NLO PDF's \cite{cteq},
$\mu=\sqrt{m^2_{\tilde{g}} + p^2_{T_{\tilde{g}}}}$, a two-loop $\alpha_s$, and 
multiply by 1.9, the ratio of inclusive NLO to LO cross sections~\cite{prospino}.

A relatively light gluino is necessary in order to obtain a bottom-quark
cross section comparable in magnitude to the pure QCD component.  Values of
$m_{\tilde g} \simeq$ 12 to 16 GeV are chosen because the resulting $\tilde
g$ decays produce $p_{Tb}$ spectra that are enhanced primarily in the
neighborhood of $p_{Tb}^{\rm min} \simeq m_{\tilde g}$ where the data show
the most prominent enhancement above the QCD expectation.  Larger values of
$m_{\tilde g}$ yield too little cross section to be of interest, and
smaller values produce more cross section than seems tolerated by the ratio
of like-sign to opposite-sign leptons from $b$ decay, as discussed below.
The choice of $m_{\tilde b}$ has an impact on the kinematics of the $b$.
After selections on $p_{Tb}^{\rm min}$, large values of $m_{\tilde b}$
reduce the cross section and, in addition, lead to shapes of the $p_{Tb}$
distribution that agree less well with the data.  The values of $m_{\tilde
b}$ and $m_{\tilde g}$ are correlated; similar results to those shown in
Fig.~1 can be obtained with $m_{\tilde g} \simeq$ 12 GeV, but $m_{\tilde b}
\simeq m_b$.
\begin{figure}[ht]
\vspace*{-0.4cm}
\begin{center}
\psfig{figure=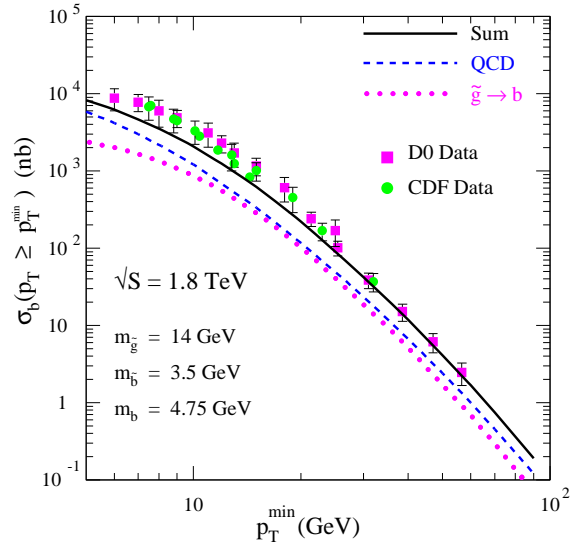,width=3.4in}
\end{center}
\vspace*{-0.3cm}
\caption{Bottom-quark cross section in $p\bar p$ collisions at $\sqrt{S}
=1.8$ TeV for $p_{Tb}>p_{Tb}^{\rm min}$ with a gluino of mass
$m_{\tilde{g}} = 14$ GeV and a bottom squark of mass $m_{\tilde{b}} = 3.5$
GeV.  The dashed curve is the central value of the NLO QCD prediction. The
dotted curve shows the $p_T$ spectrum of the $b$ from the supersymmetry
(SUSY) processes.  The solid curve is the sum of the QCD and SUSY
components.  Data are from Ref.~1.}
\label{fig:fig1}
\end{figure}

After the contributions of the NLO QCD and SUSY components are added, 
the magnitude of the bottom-quark cross section and the
shape of the integrated $p^{\rm min}_{Tb}$ distribution are described well.
A theoretical uncertainty of roughly $\pm 30\%$ may be assigned to the
final solid curve, associated with variation of the $b$ mass, the 
renormalization/factorization scale, and the parton distributions.
The SUSY process produces bottom quarks in a four-body final state and thus
their momentum correlations are different from those of QCD.  Angular
correlations between muons that arise from decays of $b$'s have been
measured \cite{cdfmix,muonexp}.  Examining the angular correlations between
$b$'s in the SUSY case, we find they are nearly indistinguishable from those
of QCD once experimental cuts are applied.
\section{Same-sign to Opposite-sign Leptons}
Since the $\tilde g$ is a Majorana particle, its decay yields both quarks
and antiquarks.  Gluino pair production and subsequent decay to $b$'s will
generate $b b$ and $\bar b \bar b$ pairs, as well as the $b \bar b$
final states that appear in QCD production.  When a gluino is highly 
relativistic, its helicity is nearly the same as its chirality.  Therefore, 
selection of $\tilde g$'s whose transverse momentum is greater than their mass 
will reduce the number of like-sign $b$'s.  In the intermediate $p_T$ region,
however, the like-sign suppression is reduced.  The cuts chosen in current 
hadron collider experiments for measurement of the ratio of like-sign to 
opposite-sign muons result in primarily unpolarized $\tilde g$'s, and, 
independent of the
$\tilde b$ mixing angle, an equal number of like-sign and opposite-sign
$b$'s is expected at production.  Our SUSY mechanism leads therefore to an 
increase of like-sign leptons in the final state after semi-leptonic decays of 
the $b$ and $\bar b$ quarks.  This increase could be confused with an enhanced 
rate of $B^0-\bar B^0$ mixing.  

Time-integrated mixing analyses of lepton pairs 
observed at hadron colliders are interpreted in terms of the 
quantity $\bar{\chi} = f_d \chi_d + f_s \chi_s$, where 
$f_d$ and $f_s$ are the fractions of
$B^0_d$ and $B^0_s$ hadrons in the sample of semi-leptonic $B$
decays, and $\chi_f$ is the time-integrated mixing probability for $B^0_f$.
Conventional $b\bar b$ pair production determines the quantity 
$LS_c = 2\bar{\chi} (1-\bar{\chi})$, the fraction of $b\bar b$ pairs that 
decay into like-sign leptons.  Our SUSY mechanism leads to a new 
expression  
\begin{equation}
LS =\frac{1}{2} \frac{\sigma_{\tilde{g}\tilde{g}}}{\sigma_{\tilde{g}\tilde{g}}+
\sigma_{\rm{qcd}}} + 
LS_c \frac{\sigma_{\rm {qcd}}}{\sigma_{\tilde{g}\tilde{g}}+\sigma_{\rm {qcd}}} = 
2 \bar{\chi}_{\rm {eff}} (1 - \bar{\chi}_{\rm {eff}}).  
\end{equation}
The factor $1/2$ arises because $N(bb + \bar{b}\bar{b}) \simeq N(b \bar{b})$ in the 
new mechanism.   
Introducing $G = \sigma_{\tilde{g}\tilde{g}} / \sigma_{\rm{qcd}}$, 
the ratio of SUSY and QCD bottom-quark cross sections after cuts, and solving for 
the effective mixing parameter, we obtain:
\begin{equation}
\bar{\chi}_{\rm {eff}}=\frac{\bar{\chi}}{\sqrt{1+G}} +{1\over 2}\left[1-
	\frac{1}{\sqrt{1+G}}\right] .
\end{equation}
We interpret the CDF collaboration's measurement as a determination of 
$\bar{\chi}_{\rm {eff}} = 0.131 \pm 0.02 \pm 0.016$~\cite{cdfmix} and note 
that it is marginally larger than the world average value 
$\bar{\chi} = 0.118 \pm 0.005$ \cite{pdg}.  

To estimate our theoretical $\bar{\chi}_{\rm {eff}}$, we assume that 
$\bar{\chi} = 0.118 \pm 0.005$ \cite{pdg} represents the contribution from 
only the QCD $b\bar{b}$ component.  We determine the ratio $G$ in the region of 
phase space where the measurement is made~\cite{cdfmix}, with both final $b$'s 
having $p_{Tb} \ge 6.5$ GeV and rapidity $| y_b | \leq 1$.  For gluino masses 
of $m_{\tilde g} =$ 14 and 16 GeV, we obtain $G =$ 0.37 and 0.28, respectively, 
with $m_{\tilde b} =$ 3.5 GeV.  We compute 
$\bar{\chi}_{\rm {eff}} = 0.17 $ for $m_{\tilde g} =$ 14 GeV, and 
$\bar{\chi}_{\rm {eff}} = 0.16 $ with $m_{\tilde g} =$ 16 GeV.
To estimate the uncertainty on $G$, we vary the renormalization/factorization 
scale at which the cross
sections are evaluated between $\mu=m_x/2$ and $\mu=2 m_x$.  Uncertainties
of $\pm 50\%$ are obtained and lead to uncertainties in $\bar{\chi}_{\rm {eff}}$ of 
$\delta\bar{\chi}_{\rm {eff}}\simeq\pm 0.02$.  Additional theoretical uncertainties
arise because there is no fully differential NLO calculation of gluino
production and subsequent decay to $b$'s.

Comparing our expectations with the CDF value, we conclude that values of 
$m_{\tilde g} > 12$ GeV lead to a calculated $\bar{\chi}_{\rm {eff}}$ that is 
consistent with the data within experimental and theoretical uncertainties.  
With $\sigma_{\tilde{g}\tilde{g}} / \sigma_{\rm{qcd}} \sim 1/3$, we can satisfy 
the magnitude and $p_T$ dependence of the $b$ production cross section and 
the mixing data.  
\section{Implications and Remarks}
\subsection{Hadron Reactions}
Among the predictions of this SUSY scenario, the most clearcut is pair
production of like-sign charged $B$ mesons at hadron colliders, $B^+B^+$ 
and $B^-B^-$.  To verify the underlying premise of this work, that the cross 
section exceeds expectations of conventional perturbative QCD, a new measurement 
of the absolute rate for $b$ production in run II of the Tevatron is important.   
A very precise measurement of $\bar{\chi}$ in run II is obviously desirable.  
Since the fraction of $b$'s from gluinos 
changes with $p_{Tb}$, we also expect a change of $\bar{\chi}$ with the cut 
on $p_{Tb}$.  The $b$ jet from $\tilde{g}$ decay into $b\widetilde{b}$ will 
contain the $\widetilde{b}$, implying unusual material associated with the 
$\widetilde{b}$ in some fraction of the $b\bar{b}$ data sample.  The existence 
of light $\widetilde b$'s means that they will be pair-produced in partonic 
processes, leading to a slight increase ($\sim 1$\%) 
in the hadronic dijet rate.  Our approach increases the $b$ production rate
at HERA and in $\gamma \gamma$ collisions at LEP by a small amount, not
enough perhaps if early experimental indications in these cases are
confirmed~\cite{herab,lepb}, but a full NLO study should be undertaken and 
better parton densities of photons are needed.
\subsection{Running of $\alpha_s$}
In the standard model, a global fit to all observables provides an indirect 
measurement of the strong coupling strength 
$\alpha_s$ at the scale of the $Z$ boson mass $M_Z$.  The value 
$\alpha_s(M_Z) \simeq 0.119 \pm 0.006$ describes most observables 
properly~\cite{alfs}.  A light $\tilde g$ with mass about 15 GeV and a light 
$\tilde b$ slow the running of $\alpha_s$ and modify $\alpha_s(M_Z)$, determined 
by extrapolation from experiments performed at energies lower than 
$m_{\tilde g}$.  The result is a shift of 0.007 in
$\alpha_s$ derived from these experiments, to $\alpha_s(M_Z) \simeq 0.125$, at the 
upper edge of the uncertainty band.  Greater precision on the determination of 
$\alpha_s(M_Z)$ from evolution would be valuable.  
Slower running of $\alpha_s(Q)$ also means a 
slower evolution of parton densities at small $x$, an effect that might be 
seen in HERA data for $Q > m_{\tilde{g}}$.  Presence of a scalar $\tilde{b}$ in 
the proton breaks the Callan-Gross relation and yields a non-zero leading-twist 
longitudinal structure function $F_L(x,Q)$ at leading-order.
\subsection{$\widetilde b$-onia}
Possible bound states of bottom squark pairs could be seen as $J^P = 0^+$, 
$1^-$, $2^+$, ... mesonic resonances in $\gamma \gamma$ reactions and in 
$p\bar{p}$ formation, with masses in the 4 to 10 GeV range.  They could 
show up as narrow states in the $\mu^+ \mu^-$ invariant mass spectra at hadron 
colliders, between the $J/\Psi$ and $\Upsilon$.  At an $e^+ e^-$ collider, the 
intermediate photon requires production of a $J^{PC} = 1^{--}$ state.  Bound 
states of low mass squarks with charge $2/3$ were studied with a potential 
model~\cite{Nappi}.  The small leptonic widths were found to preclude bounds 
for $m_{\widetilde{q}} > 3$ GeV.  For bottom squarks with charge $-1/3$, the 
situation is more difficult. 
\subsection{$\widetilde b$ lifetime and observability}
Strict R-parity conservation in the MSSM forbids $\widetilde{b}$ decay 
unless there is a lighter supersymmetric particle.  R-parity-violating and 
lepton-number-violating decay of the $\widetilde{b}$ into at least one lepton is 
disfavored by the CLEO data~\cite{CLEO} and would imply the presence of an extra 
lepton, albeit soft, in some fraction of $b$ jets observed at hadron colliders.  
The baryon-number-violating R-parity-violating (${\not \! R_p}$) term in the MSSM 
superpotential is 
${\cal W}_{\not \! R_p} = \lambda_{ijk}^{\prime\prime}U_i^cD_j^cD_k^c$; 
$U^c_i$ and $D^c_i$ are right-handed-quark singlet chiral
superfields; and $i,j,k$ are generation indices.  The limits on individual 
${\not \! R_p}$ and baryon-number violating couplings 
$\lambda''$ are relatively weak for third-generation squarks~\cite{Allanach,BHS}, 
$\lambda''_{ijk} < 0.5$ to $1$.  

The possible ${\not \! R_p}$ decay channels 
for the $\widetilde{b}$ are $123: \bar{\widetilde{b}} \rightarrow u+s$; 
$213: \bar{\widetilde{b}} \rightarrow c+d$; and 
$223: \bar{\widetilde{b}} \rightarrow c+s$.  The hadronic width is~\cite{BHS} 
\bea
\Gamma(\widetilde b \rightarrow \rm{jet} + \rm{jet}) =
\frac{m_{\widetilde b}}{2\pi} \sin^2\theta_{\widetilde{b}} 
\sum_{j<k} |\lambda^{\prime\prime}_{ij3}|^2 .
\eea
If $m_{\tilde b} = 3.5$ GeV, $\Gamma(\widetilde b \rightarrow i j) = 
0.08 |\lambda^{\prime\prime}_{ij3}|^2$ GeV.  Unless all 
$\lambda^{\prime\prime}_{ij3}$ are extremely small, the $\widetilde {b}$ will 
decay quickly and leave soft jets in the cone around the $b$.  $b$-jets with an 
extra $c$ are possibly disfavored by CDF, but a detailed simulation is needed.

If the $\widetilde{b}$ is relatively stable, the $\widetilde{b}$ could pick 
up a light $\bar{u}$ or $\bar{d}$ and become a $\widetilde{B}^-$ or 
$\widetilde{B}^0$ ``mesino" with $J = 1/2$, the superpartner of the $B$ meson.  
The mass of the mesino would fall roughly in the range $3$ to $7$ GeV for 
the interval of $\widetilde{b}$ masses we consider.  The charged mesino could fake 
a heavy muon if its hadronic cross 
section is small and if it survives passage through the hadron calorimeter and exits 
the muon chambers.  Extra muon-like tracks would then appear in a fraction of the 
$b \bar{b}$ event sample, but tracks that left some activity in the hadron calorimeter.  
The mesino has baryon number zero but acts like a 
heavy $\bar{p}$ -- perhaps detectable with a time-of-flight apparatus. 
A long-lived $\widetilde b$ is not excluded by conventional searches at hadron 
and lepton colliders, but an analysis~\cite{baeretal} similar to that for 
$\tilde{g}$'s  should be done to verify that there are no 
additional constraints on the allowed range of $\tilde b$ masses and lifetimes.
\vspace*{-0.2cm}
\section*{Acknowledgments}
\vspace*{-0.2cm}
I am indebted to Brian~Harris, David~E.~Kaplan, Zack~Sullivan, Tim~Tait, and Carlos~
Wagner for their collaboration and excellent suggestions.  I have benefitted from 
valuable discussions with Barry Wicklund, Tom LeCompte, and Harry Lipkin.  I 
congratulate Tony Sanda and his colleagues for organizing an excellent Fourth 
International Workshop on B Physics and CP Violation in an exceptional location.  
This research was supported by the U.S. Department of Energy under Contract 
W-31-109-ENG-38.

\end{document}